\documentclass[twocolumn,superscriptaddress]{revtex4-2}

\usepackage{graphicx}
\usepackage{dcolumn}
\usepackage{bm}

\usepackage{color}
\usepackage{amsmath}
\usepackage{float}
\usepackage{booktabs}
\usepackage{caption}
\usepackage{graphicx}
\usepackage{ragged2e}  
\DeclareUnicodeCharacter{2550}{\ensuremath{\equiv}}
\usepackage{hyperref}
\definecolor{darkred}{rgb}{0.25,0,0}
\definecolor{darkgreen}{rgb}{0,0.25,0}
\definecolor{darkblue}{rgb}{0,0,1}
\hypersetup{colorlinks,linkcolor=darkblue,filecolor=black,urlcolor=darkblue,citecolor=darkblue}
\usepackage{multirow}
\usepackage{textcomp}
\usepackage[compact]{titlesec}
\vspace{-0.5cm} 
\begin{document}

\preprint{APS/123-QED}

\title{Revisiting phonon thermal transport in two-dimensional gallium nitride: higher-order phonon-phonon and phonon-electron scattering}


\author{Jianshi Sun}
\affiliation{Institute of Micro/Nano Electromechanical System, College of Mechanical Engineering, Donghua University, Shanghai 201620, China
}%

\author{Xiangjun Liu}
\affiliation{%
 Institute of Micro/Nano Electromechanical System, College of Mechanical Engineering, Donghua University, Shanghai 201620, China
}%

\author{Yucheng Xiong}
\affiliation{%
Institute of Micro/Nano Electromechanical System, College of Mechanical Engineering, Donghua University, Shanghai 201620, China
}

\author{Yuhang Yao}
\affiliation{%
Institute of Micro/Nano Electromechanical System, College of Mechanical Engineering, Donghua University, Shanghai 201620, China
}

\author{Xiaolong Yang}
\affiliation{%
 College of Physics and Center of Quantum Materials \& Devices, Chongqing University, Chongqing 401331, China
}%

\author{Cheng Shao}
\affiliation{
 Thermal Science Research Center, Shandong Institute of Advanced Technology, Jinan, Shandong 250103, China
}

\author{Shouhang Li}
\email{shouhang.li@dhu.edu.cn}
\affiliation{Institute of Micro/Nano Electromechanical System, College of Mechanical Engineering, Donghua University, Shanghai 201620, China
}%

\date{\today}

\begin{abstract}
Two-dimensional gallium nitride (2D-GaN) has great potential in power electronics and optoelectronics. Heat dissipation is a critical issue for these applications of 2D-GaN. Previous studies showed that higher-order phonon-phonon scattering has extremely strong effects on the lattice thermal conductivity ($\kappa_{\text {lat}}$) of 2D-GaN, which exhibits noticeable discrepancies with $\kappa_{\text {lat}}$ calculated from molecular dynamics. In this work, it is found that the fourth-order interatomic force constants (4th-IFCs) of 2D-GaN are quite sensitive to atomic displacement in the finite different method. The effects of the four-phonon scattering can be severely overestimated with non-convergent 4th-IFCs. The $\kappa_{\text {lat}}$ from three-phonon scattering is reduced by 65.6\% due to four-phonon scattering. The reflection symmetry allows significantly more four-phonon processes than three-phonon processes. It was previously thought the electron-phonon interactions have significant effects on the $\kappa_{\text {lat}}$ of two-dimensional materials. However, the effects of phonon-electron interactions on the $\kappa_{\text {lat}}$ of both \textit{n}-type and \textit{p}-type 2D-GaN at high charge carrier concentrations can be neglected due to the few phonon-electron scattering channels and the relatively strong four-phonon scattering.
\end{abstract}

\maketitle

2D-GaN has attracted great attention in power electronics\cite{teo2020recent} and optoelectronics\cite{yeh2012vertical} devices due to its large surface-to-volume ratios, structural stability, and ease of forming heterostructures. The 2D-GaN single crystal has been successfully synthesized by chemical vapor deposition with excellent electrostatic gate control performances, an ON/OFF ratio exceeding $10^6$, and mobility up to 160 $\, \text{cm}^2 \, \text{V}^{-1} \, \text{s}^{-1}$\cite{chen2018growth}. Efficient heat dissipation is crucial for the reliable performance and stable functionality of devices based on 2D-GaN, and a high thermal conductivity of 2D-GaN is desired\cite{chen2019gan}. Due to the challenges in preparing experimental samples and conducting thermal measurements, the investigation of the thermal transport in 2D-GaN mainly depends on theoretical calculations. Especially, the method of extracting force constants from first-principles calculations combined with the Peierls-Boltzmann transport equation (PBTE) has shown to be robust in predicting the $\kappa_{\text {lat}}$ of 2D materials\cite{yue2019controlling,tong2021ultralow,yang2021tuning,liu2020strong,guo2024four,huang2019abnormally}. In addition, PBTE can provide abundant physical information, like the phonon linewidth, phonon mean free path, Grüneisen parameter, \textit{etc}., which are critical to understanding the thermal transport mechanisms in 2D materials.

There have been a few theoretical studies on the $\kappa_{\text {lat}}$ of 2D-GaN based on PBTE\cite{jiang2017phonon,qin2017anomalously,qin2017orbitally,shen2022two,cai2021effect}. However, there exist great discrepancies in the reported values. The room-temperature $\kappa_{\text {lat}}$ of 2D-GaN from Jiang \textit{et al.}\cite{jiang2017phonon} and Qin \textit{et al.}\cite{qin2017orbitally} are 37 W/mK and 15 W/mK, respectively, even the same methodology was been used. The differences may be attributed to the different exchange-correlation (XC) functionals, pseudopotentials, and sampling grids of the first Brillouin zone (BZ). Recent studies found that the higher-order phonon-phonon (four-phonon) scattering can greatly impede phonon thermal transport in 2D materials\cite{tong2021ultralow,guo2024four,sun2023four,feng2018four,han2022raman,li2024convergent}. Notably, Sun \textit{et al.} reported that the $\kappa_{\text {lat}}$ of 2D-GaN decreases from 21.9 to 1.39 W/mK due to four-phonon scattering, which is more than an order of magnitude reduction\cite{sun2023four}. This work also represents the most significant reduction in the $\kappa_{\text {lat}}$ aroused by four-phonon scattering up to date. However, Karaaslan \textit{et al.} reported the $\kappa_{\text {lat}}$ of 2D-GaN to be 15 W/mK via classical molecular dynamics (MD), where the four-phonon scatterings are naturally captured\cite{karaaslan2020assessment}. Therefore, the controversies regarding the precise value of $\kappa_{\text {lat}}$ of 2D-GaN are still open. Recently, Zhou \textit{et al.} found that the 4th-IFCs are sensitive to the atomic displacement (\textit{h}) when calculated using the finite difference method\cite{zhou2023extreme}. For bulk silicon, the widely used value of \textit{h} = 0.01 Å in calculating 4th-IFCs can even result in an ultralow $\kappa_{\text {lat}}$ of $\sim$20 W/mK at room temperature. The convergence issue of \textit{h} is also crucial in calculating the third-order interatomic force constants (3rd-IFCs) of GaP. The $\kappa_{\text {lat}}$ is increased by $\sim$47\% when \textit{h} increases from 0.01 Å to the convergent value of 0.03 Å\cite{dongre2022thermal}. A similar trend was also observed in the strong anharmonicity crystal TI$_3$VSe$_4$ and silver halide\cite{li2023first,ouyang2023role}. Therefore, the extremely strong influence of four-phonon scattering on the $\kappa_{\text {lat}}$ of 2D-GaN may be attributed to the improper 4th-IFCs used in PBTE waiting to be verified.

On the other hand, as the channel layer of power electronics, a large electrical conductivity is highly desired in 2D-GaN. This is often achieved by heavy doping in 2D-GaN, which can have a negative influence on its thermal transport. An additional term, namely phonon-electron scattering, is introduced when there are plenty of charge carriers. The phonon-electron scattering was found to significantly decrease the $\kappa_{\text {lat}}$ of 2D materials due to the weakened effect of three-phonon scattering caused by reflection symmetry in previous studies\cite{yue2019controlling,liu2020strong,wang2022roles}. For instance, the $\kappa_{\text {lat}}$ of silicene is reduced by more than 40\% when the hole concentration is increased to $10^{13} \, \text{cm}^{-2}$\cite{yue2019controlling}. Similar trends were also observed in other 2D semiconductor materials, such as MoS$_2$ and PtSSe\cite{liu2020strong}. It remains unknown whether phonon-electron interactions have effects on $\kappa_{\text {lat}}$ of 2D-GaN.

In this work, we revisit the lattice thermal conductivity of 2D-GaN by rigorously considering four-phonon and phonon-electron scattering. For the first time, we found that 4th-IFCs are quite sensitive to atomic displacement in 2D material and the surface roughness of exchange-correlation functional. The widely used 0.01 Å of the atomic displacement in calculating 4th-IFCs can arouse a dramatic reduction in $\kappa_{\text {lat}}$. The 4th-IFCs can only be converged when the atomic displacement is equal to or larger than 0.02 Å. With the converged 4th-IFCs, the $\kappa_{\text {lat}}$ from PBTE can match the results from molecular dynamics well. The impact of phonon-electron interactions on the $\kappa_{\text {lat}}$ can be ignored in both \textit{n}-type and \textit{p}-type 2D-GaN at high charge carrier concentrations since small electron density of states within the Fermi window and four-phonon scattering is dominant among all kinds of phonon scattering terms.  

The first-principles calculations are performed using the Quantum Espresso package\cite{giannozzi2009quantum}. We use local density approximation (LDA) electron XC functional\cite{jackson1990accurate} and optimized relativistic norm-conserving pseudopotentials\cite{hamann2013optimized} from PseudoDojo\cite{van2018pseudodojo}. The lattice vectors and atomic positions are fully relaxed based on the Broyden-Fretcher-Goldfarb-Shanno (BFGS) optimization method\cite{fletcher1970new,broyden1970convergence,goldfarb1970family,shanno1970conditioning} and the convergence threshold of both energy and force are set to $10^{-8}$ Ry. The kinetic energy cutoff for plane waves is set to be 120 Ry. The Brillouin zone is sampled with a 12×12×1 Monkhorst-Pack $\mathbf{k}$-point mesh to ensure convergence. The threshold of electron total energy is set to be $10^{-10}$ Ry. The spin-orbit coupling effects are not included, as the changes in the electron band structure near the conduction band minimum (CBM) and valence band maximum (VBM) are tiny, as shown in Figure S1, Supplemental Material. The optimized in-plane lattice constant is 3.16 Å for 2D-GaN, which agrees well with the experimental value of 3.20 Å\cite{al2016two}. A vacuum layer of 15 Å is introduced in the out-of-plane direction to eliminate the interactions between different layers. The harmonic force constants are calculated from density-functional perturbation theory\cite{fugallo2013ab,baroni2001phonons}. The $\mathbf{q}$-point mesh is set to be 6×6×1, and the energy threshold is $10^{-17}$ Ry to guarantee convergence. Also, the dielectric constant and Born effective charge are calculated to account for the long-range electrostatic interactions. The convergence of the phonon dispersion with respect to the $\mathbf{q}$-point meshes is shown in Figure S2, Supplemental Material. The 3rd-IFCs are calculated from the finite difference method\cite{li2014shengbte}. We employ a 10×10×1 supercell and up to the eighth nearest neighbors in calculating 3rd-IFCs. As shown in Figure S3, Supplemental Material, an atomic displacement of \textit{h} = 0.01 Å is enough to achieve convergent for 3rd-IFCs. The 4th-IFCs and four-phonon scattering rate are calculated with the FourPhonon package\cite{han2022fourphonon}. We employ a 5×5×1 supercell and up to the third nearest neighbors in calculating 4th-IFCs. The convergence tests of the $\kappa_{\text {lat}}$ respect to the cutoff distances, $\mathbf{q}$-point meshes, and scalebroads in the delta function are shown in Figure S4, Supplemental Material. For the calculation of phonon-electron scattering rate, maximally-localized Wannier functions\cite{marzari2012maximally} are used in the interpolation of electron-phonon matrix elements. The electron-phonon coupling matrix elements are first calculated under coarse $\mathbf{k}$/$\mathbf{q}$-point meshes (12×12×1/6×6×1) and then interpolated to dense $\mathbf{k}$/$\mathbf{q}$-point meshes (60×60×1/48×48×1). The in-house modified D3Q package\cite{fugallo2013ab} is employed to calculate $\kappa_{\text {lat}}$ with four-phonon and phonon-electron scatterings included. The phonon-isotope scattering is incorporated in all calculations\cite{tamura1983isotope,sun2023light}. $\kappa_{\text {lat}}$ is converged with respect to a $\mathbf{q}$-point mesh of 48×48×1. The thickness of the 2D-GaN is determined by the van der Waals diameter of the gallium atom, which is 3.74 Å. The rigid shift of the Fermi energy is employed to imitate the change of carrier concentration \cite{li2019resolving}. The expressions for lattice thermal conductivity and all involved scattering rates in this work are provided in Section I of the Supplemental Material.

\begin{figure}[H]
    \centering
    \includegraphics[width=0.95\columnwidth]{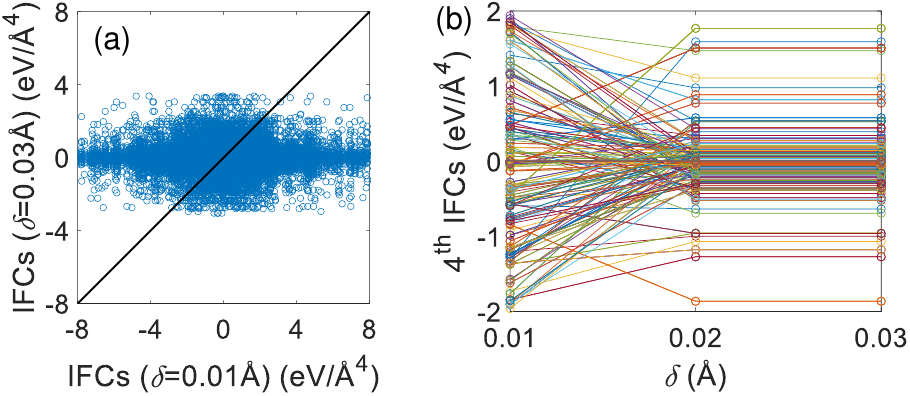}
    \caption{(a) The comparison of 4th-IFCs calculated by using finite difference method with \textit{h} of 0.01 and 0.03 Å. (b) The convergence test of \textit{h} for 4th-IFCs with respect to \textit{h}.}
    \label{fig: Figure 1}
\end{figure}

For phonon-phonon scattering rate calculation, the procedures to achieve convergent three-phonon scattering rate are matured. Therefore, we focus on the accuracy of the four-phonon scattering rate in this work. Figure \ref{fig: Figure 1}(a) shows the 4th-IFCs of 2D-GaN calculated using the finite difference method with \textit{h} values of 0.01 and 0.03 Å. Note that we only show 4th-IFCs within the range of -8 to 8 eV/ Å$^{-4}$, as this range covers most of 4th-IFCs components (93\%). Besides, these smaller 4th-IFCs have a significant influence on $\kappa_{\text {lat}}$\cite{zhou2023extreme}. The 4th-IFCs corresponding to commonly used \textit{h} = 0.01 Å\cite{han2022fourphonon} are much scattered compared to those corresponding to \textit{h} = 0.03 Å. We further conduct the convergence test on \textit{h}, as shown in Figure \ref{fig: Figure 1}(b). The 4th-IFCs are converged when \textit{h} is greater than or equal to 0.02 Å. This indicates that the relatively small \textit{h} leads to non-smooth characteristics in the fourth-order derivatives of the potential energy surface. Therefore, a displacement exceeding 0.02 Å is required to overcome the roughness of the potential energy surface. A similar trend is also observed in three-dimensional bulk materials, like Si, BAs, and NaCl\cite{zhou2023extreme}. Therefore, we adopt \textit{h} = 0.03 Å in the subsequent calculations for 4th-IFCs.

\begin{figure}[H]
    \centering
    \includegraphics[width=0.95\columnwidth]{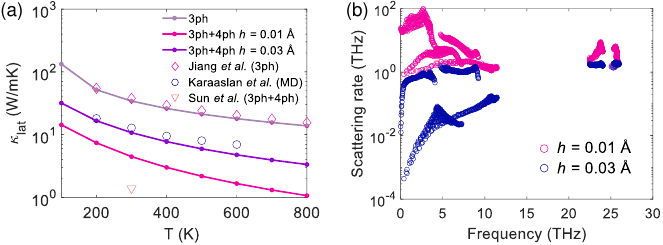}
    \captionsetup{justification=justified}
    \caption{(a) The $\kappa_{\text {lat,3}}$ and $\kappa_{\text {lat,3+4}}$ of 2D-GaN with respect to \textit{h} = 0.01 and 0.03 Å. The scattered points are the results reported by Jiang \textit{et al.}\cite{jiang2017phonon} (diamond) and Sun \textit{et al.}\cite{sun2023four} (triangle) from PBTE, and Karaaslan \textit{et al.}\cite{karaaslan2020assessment} (circle) from MD, respectively. (b) The four-phonon scattering rates are calculated with \textit{h} = 0.01 and 0.03 Å at room temperature.}
    \label{fig: Figure 2}
\end{figure}

Figure \ref{fig: Figure 2}(a) shows the lattice thermal conductivity calculated from PBTE. The three-phonon lattice thermal conductivity ($\kappa_{\text {lat,3}}$) is 36 W/mK at room temperature, which agrees well with previous results from PBTE\cite{jiang2017phonon,kocabacs2023gaussian}. However, this value is much larger than the results from MD\cite{karaaslan2020assessment}. With four-phonon scattering further included, the lattice thermal conductivity ($\kappa_{\text {lat,3+4}}$) is reduced by 87.7\% compared to $\kappa_{\text {lat,3}}$ with the 4th-IFCs calculated with \textit{h} = 0.01 Å. Note that a significant reduction is also observed in Ref. \cite{sun2023four}. However, Ref. \cite{sun2023four} reports much lower $\kappa_{\text {lat,3+4}}$ than ours, which can be attributed to the difference in pseudopotentials. Interestingly, $\kappa_{\text {lat,3+4}}$ calculated with \textit{h} = 0.01 Å is significantly smaller than the result from MD\cite{karaaslan2020assessment}. Therefore, we recalculate $\kappa_{\text {lat,3+4}}$ with the 4th-IFCs calculated with \textit{h} = 0.03 Å, which can ensure the convergence of the 4th-IFCs. The $\kappa_{\text {lat,3+4}}$ matches well with the results reported from MD \cite{karaaslan2020assessment} for the entire temperature range when using \textit{h} = 0.03 Å. Note that we also calculate the $\kappa_{\text {lat,3+4}}$ of 2D-GaN using the PBEsol electron XC functional. It is found that the $\kappa_{\text {lat,3+4}}$ calculated with \textit{h} = 0.01 Å matches well with the results reported in Ref. \cite{sun2023four}. However, the $\kappa_{\text {lat,3}}$ is reduced by only 62\% when using \textit{h} = 0.03 Å, contradicting the reported 94\% reduction in Ref. \cite{sun2023four}. It is indicated that the 4th-IFCs are sensitive to \textit{h} rather than the pseudopotentials. To further quantitatively analyze the effects of \textit{h} on $\kappa_{\text {lat,3+4}}$, we examine four-phonon scattering rates calculated with different \textit{h} values. As shown in Figure \ref{fig: Figure 2}(b), the four-phonon scattering rates of phonons with frequencies below 12 THz are notably overestimated when using the improper \textit{h} value of 0.01 Å. This leads to a significant underestimation of $\kappa_{\text {lat,3+4}}$ discussed above. 

As shown in Table \ref{Table 1}, the $\kappa_{\text {lat,3+4}}$ is mainly contributed by the acoustic phonon branches. There is a redistribution in the contributions of phonon branches to $\kappa_{\text {lat}}$ with different \textit{h} values. Especially, the change in the ZA branch contribution is more than 10 times. Given that 0.01 Å is the default value for \textit{h} in the Fourthorder script\cite{han2022fourphonon} for extracting 4th-IFCs, the reports on the effects of four-phonon scattering on $\kappa_{\text {lat}}$ of 2D materials in most literature need to be further validated. Furthermore, 2D GaN exhibits reflection symmetry in the out-of-plane direction. Due to this symmetry, phonon scattering processes involving an odd number of ZA phonons are prohibited, greatly suppressing three-phonon scatterings with one or three ZA phonons. However, with the four-phonon scattering further included, the scattering process in which two ZA phonons redistribute into two other ZA phonons is significantly enhanced.  The contribution of ZA modes to the $\kappa_{\text {lat}, 3}$ decreases by more than threefold when the four-phonon scattering is incorporated, as shown in Table \ref{Table 1}. Therefore, four-phonon scattering in 2D-GaN has a significant impact on $\kappa_{\text {lat}}$.

Note that $\kappa_{\text {lat,3+4}}$ reported here is obtained by a mixing scheme to solve the PBTE where the three-phonon scattering rates are treated iteratively while the relaxation time approximation is adopted for the four-phonon scattering rates. Recently, Han \textit{et al.} found that the full iterative solution of the PBTE can have significant effects on the $\kappa_{\text {lat}}$ of graphene\cite{han2023thermal}. We further evaluate the four-phonon scattering rate corresponding to normal ($1 / \tau_{\lambda}^{\mathrm{4ph(N)}}$) and Umklapp ($1 / \tau_{\lambda}^{\mathrm{4ph(U)}}$) processes, as shown in Figure S5 (a), Supplemental Material. It is found that $1 / \tau_{\lambda}^{\mathrm{4ph(N)}}$ is larger than $1 / \tau_{\lambda}^{\mathrm{4ph(U)}}$. To further analyze the effects of the four-phonon iterative solution on $\kappa_{\text {lat}}$, we calculate $\kappa_{\text {lat}}$ as a function of temperature for three-phonon (RTA), three-phonon (iterative), three-phonon (RTA) + four-phonon (RTA), and three-phonon (iterative) + four-phonon (RTA). As shown in Figure S5 (b), Supplemental Material, the iterative solutions play a crucial role in determining the $\kappa_{\text {lat,3}}$. However, with the four-phonon RTA scattering rate further included, the effect of iterative solutions becomes negligible. Therefore, it can be expected the four-phonon iterative solution can have a weak effect on the $\kappa_{\text {lat}}$ of 2D-GaN.

\begin{table}[H]
  \centering
  \caption{The contributions to $\kappa_{\text {lat}}$ concerning \textit{h} = 0.01 and 0.03 Å and only considering three phonon-phonon scattering rates (3ph) of the out-of-plane acoustic (ZA), in-plane transverse acoustic (TA), in-plane longitudinal acoustic (LA), out-of-plane optical (ZO), in-plane transverse optical (TO), and in-plane longitudinal optical (LO) phonon branches.}
  \begin{tabular}{lcccccc}
    \hline
    Mode & ZA & TA & LA & ZO & TO & LO \\
    \hline
    Ratio (\textit{h} = 0.01 Å) & 0.4\% & 42.7\% & 44.8\% & 11.6\% & 0.3\% & 0.2\% \\
    \hline
    Ratio (\textit{h} = 0.03 Å) & 7.3\% & 59.2\% & 27.6\% & 5.6\% & 0.2\% & 0.1\% \\
    \hline
    Ratio (3ph) & 25.9\% & 39.8\% & 17.6\% & 3.5\% & 10.3\% & 2.9\% \\
    \hline
  \end{tabular}
  \label{Table 1}
\end{table}

\begin{figure}[H]
    \centering
    \includegraphics[width=1.00\columnwidth]{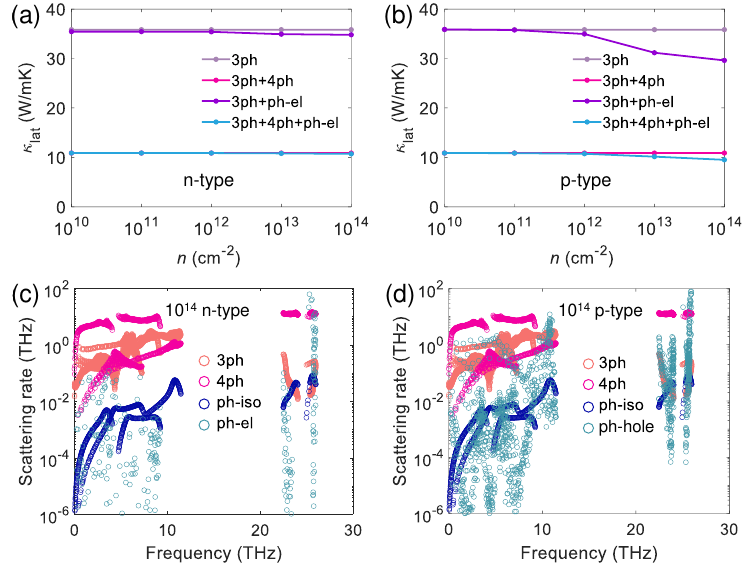}
    \caption{The lattice thermal conductivity as a function of charge carrier concentration at room temperature for (a) \textit{n}-type and (b) \textit{p}-type 2D-GaN with different combinations of phonon scattering rates. The (c) phonon-electron scattering rate and (d) phonon-hole scattering rate compared with three-phonon, four-phonon, and phonon-isotope scattering rates at room temperature with the carrier concentration of $10^{14} \, \text{cm}^{-2}$.}
    \label{fig: Figure 3}
\end{figure}

Next, we discuss the effects of electron-phonon interactions (EPI) on the $\kappa_{\text {lat}}$ of 2D-GaN. For intrinsic 2D-GaN, the Fermi energy is far away from the CBM and VBM, resulting in a lower charge carrier concentration, as shown in Figure S6, Supplemental Material. Therefore, the impact of EPI on $\kappa_{\text {lat}}$ can be neglected. However, previous studies showed that the EPI can have significant effects on $\kappa_{\text {lat}}$ at high charge carrier concentrations in 2D materials\cite{yue2019controlling,liu2020strong}. Therefore, we further calculate the $\kappa_{\text {lat}}$ of 2D-GaN for both \textit{n}-type and \textit{p}-type at charge carrier concentrations of $10^{10}-10^{14} \, \text{cm}^{-2}$, as shown in Figure \ref{fig: Figure 3}(a) and (b). For \textit{n}-type 2D-GaN, the $\kappa_{\text {lat,3}}$ falls within the range of 35-36 W/mK with phonon-electron scattering included, which is quite close to the value of the intrinsic case. In contrast, for \textit{p}-type 2D-GaN, the reduction in  $\kappa_{\text {lat,3}}$ is relatively larger, especially at high hole concentrations. Similar phenomena were also observed in our recent investigation into three-dimensional GaN\cite{sun2024unexpected}. Furthermore, the EPI almost has no effects on $\kappa_{\text {lat,3+4}}$ of both \textit{n}-type and \textit{p}-type 2D-GaN. This is totally different from the conclusions in other 2D materials\cite{yue2019controlling,liu2020strong,guo2024four}.

The anomalous observation in EPI effects on $\kappa_{\text {lat}}$ primarily stems from phonon scattering rates in 2D-GaN. The room-temperature three-phonon ($1 / \tau_{\lambda}^{\mathrm{3ph}}$), four-phonon ($1 / \tau_{\lambda}^{\mathrm{4ph}}$), phonon-isotope  ($1 / \tau_{\lambda}^{\mathrm{ph}-\text{iso}}$), and phonon-electron/hole ($1 / \tau_{\lambda}^{\mathrm{ph}-\text{el}}$/$1 / \tau_{\lambda}^{\mathrm{ph}-\text{hole}}$) scattering rates are shown in Figure \ref{fig: Figure 3}(c) and (d).  The $1 / \tau_{\lambda}^{\mathrm{ph}-\text{el}}$ is much smaller than the $1 / \tau_{\lambda}^{\mathrm{3ph}}$ for low-frequency phonons, which have the main contribution to $\kappa_{\text {lat}}$. In contrast, the $1 / \tau_{\lambda}^{\mathrm{ph}-\text{hole}}$ of a few phonon modes below 12 THz are comparable to their $1 / \tau_{\lambda}^{\mathrm{3ph}}$. Although the $1 / \tau_{\lambda}^{\mathrm{ph}-\text{hole}}$ for phonons with frequencies larger than 20 THz can be larger than $1 / \tau_{\lambda}^{\mathrm{3ph}}$, its effects on $\kappa_{\text {lat}}$ is marginal since those high-frequency phonons have little contribution to $\kappa_{\text {lat}}$, as shown in Figure S7, Supplemental Material. The distinctions in the magnitudes of $1 / \tau_{\lambda}^{\mathrm{ph}-\text{el}}$ and $1 / \tau_{\lambda}^{\mathrm{ph}-\text{hole}}$ are primarily attributed to the differences in the electron density of states (DOS) within the Fermi window for \textit{n}-type and \textit{p}-type 2D GaN. As shown in Figure S8, Supplemental Material, the DOS for \textit{p}-type is notably larger than that for \textit{n}-type at $10^{14} \, \text{cm}^{-2}$. However, the DOS for both \textit{n}-type and \textit{p}-type 2D GaN are much smaller than those of other 2D materials like silicene, phosphorene, and graphene\cite{yue2019controlling,yang2021tuning}. Therefore, the $1 / \tau_{\lambda}^{\mathrm{ph}-\text{el}}$ and $1 / \tau_{\lambda}^{\mathrm{ph}-\text{hole}}$ of 2D-GaN are much smaller than those of other 2D materials. The $1 / \tau_{\lambda}^{\mathrm{4ph}}$ is much larger than $1 / \tau_{\lambda}^{\mathrm{ph}-\text{el}}$ and $1 / \tau_{\lambda}^{\mathrm{ph}-\text{hole}}$, which further eliminates the EPI effects on  $\kappa_{\text {lat}}$.

In summary, we perform a first-principles investigation on the phonon thermal transport in 2D-GaN, taking into account phonon anharmonicity up to the fourth order and electron-phonon interactions. It is found that the roughness of the potential energy surface and the 4th-IFCs are quite sensitive to atomic displacement, while the 3rd-IFCs are not. To accurately predict the lattice thermal conductivity of 2D materials, a rigorous convergence test on the 4th-IFCs is required. The three-phonon thermal conductivity of 2D-GaN can be reduced by 65.6\% when the four-phonon scattering is incorporated. The reduction is relatively smaller compared to the order of magnitude reduction in the previous report. The effect of phonon-electron scattering on lattice thermal conductivity of both \textit{n}-type and \textit{p}-type 2D-GaN can be neglected due to the small electron density of states within the Fermi window and the relatively strong four-phonon scattering. The conclusions in the phonon thermal transport in 2D-GaN can be further extended to other 2D semiconductor materials.

S.L. was supported by the National Natural Science Foundation of China (Grant No. 12304039) the Shanghai Municipal Natural Science Foundation (Grant No. 22YF1400100), the Fundamental Research Funds for the Central Universities (Grant No.2232022D-22), and the startup funding for youth faculty from the College of Mechanical Engineering of Donghua University. X.L. was supported by the Shanghai Municipal Natural Science Foundation (Grant No. 21TS1401500) and the National Natural Science Foundation of China (Grant No.52150610495 and 12374027)

\bibliography{bibliography}

\begin{thebibliography}{45}%
\makeatletter
\providecommand \@ifxundefined [1]{%
 \@ifx{#1\undefined}
}%
\providecommand \@ifnum [1]{%
 \ifnum #1\expandafter \@firstoftwo
 \else \expandafter \@secondoftwo
 \fi
}%
\providecommand \@ifx [1]{%
 \ifx #1\expandafter \@firstoftwo
 \else \expandafter \@secondoftwo
 \fi
}%
\providecommand \natexlab [1]{#1}%
\providecommand \enquote  [1]{``#1''}%
\providecommand \bibnamefont  [1]{#1}%
\providecommand \bibfnamefont [1]{#1}%
\providecommand \citenamefont [1]{#1}%
\providecommand \href@noop [0]{\@secondoftwo}%
\providecommand \href [0]{\begingroup \@sanitize@url \@href}%
\providecommand \@href[1]{\@@startlink{#1}\@@href}%
\providecommand \@@href[1]{\endgroup#1\@@endlink}%
\providecommand \@sanitize@url [0]{\catcode `\\12\catcode `\$12\catcode `\&12\catcode `\#12\catcode `\^12\catcode `\_12\catcode `\%12\relax}%
\providecommand \@@startlink[1]{}%
\providecommand \@@endlink[0]{}%
\providecommand \url  [0]{\begingroup\@sanitize@url \@url }%
\providecommand \@url [1]{\endgroup\@href {#1}{\urlprefix }}%
\providecommand \urlprefix  [0]{URL }%
\providecommand \Eprint [0]{\href }%
\providecommand \doibase [0]{https://doi.org/}%
\providecommand \selectlanguage [0]{\@gobble}%
\providecommand \bibinfo  [0]{\@secondoftwo}%
\providecommand \bibfield  [0]{\@secondoftwo}%
\providecommand \translation [1]{[#1]}%
\providecommand \BibitemOpen [0]{}%
\providecommand \bibitemStop [0]{}%
\providecommand \bibitemNoStop [0]{.\EOS\space}%
\providecommand \EOS [0]{\spacefactor3000\relax}%
\providecommand \BibitemShut  [1]{\csname bibitem#1\endcsname}%
\let\auto@bib@innerbib\@empty
\bibitem [{\citenamefont {Teo}\ \emph {et~al.}(2020)\citenamefont {Teo}, \citenamefont {Chowdhury}, \citenamefont {Zhang}, \citenamefont {Palacios}, \citenamefont {Yamanaka},\ and\ \citenamefont {Yamaguchi}}]{teo2020recent}%
  \BibitemOpen
  \bibfield  {author} {\bibinfo {author} {\bibfnamefont {K.~H.}\ \bibnamefont {Teo}}, \bibinfo {author} {\bibfnamefont {N.}~\bibnamefont {Chowdhury}}, \bibinfo {author} {\bibfnamefont {Y.}~\bibnamefont {Zhang}}, \bibinfo {author} {\bibfnamefont {T.}~\bibnamefont {Palacios}}, \bibinfo {author} {\bibfnamefont {K.}~\bibnamefont {Yamanaka}},\ and\ \bibinfo {author} {\bibfnamefont {Y.}~\bibnamefont {Yamaguchi}},\ }\bibfield  {title} {\bibinfo {title} {Recent development in 2d and 3d gan devices for rf and power electronics applications},\ }in\ \href@noop {} {\emph {\bibinfo {booktitle} {2020 IEEE International Symposium on Radio-Frequency Integration Technology (RFIT)}}}\ (\bibinfo {organization} {IEEE},\ \bibinfo {year} {2020})\ pp.\ \bibinfo {pages} {22--24}\BibitemShut {NoStop}%
\bibitem [{\citenamefont {Yeh}\ \emph {et~al.}(2012)\citenamefont {Yeh}, \citenamefont {Lin}, \citenamefont {Ahn}, \citenamefont {Stewart}, \citenamefont {Daniel~Dapkus},\ and\ \citenamefont {Nutt}}]{yeh2012vertical}%
  \BibitemOpen
  \bibfield  {author} {\bibinfo {author} {\bibfnamefont {T.-W.}\ \bibnamefont {Yeh}}, \bibinfo {author} {\bibfnamefont {Y.-T.}\ \bibnamefont {Lin}}, \bibinfo {author} {\bibfnamefont {B.}~\bibnamefont {Ahn}}, \bibinfo {author} {\bibfnamefont {L.~S.}\ \bibnamefont {Stewart}}, \bibinfo {author} {\bibfnamefont {P.}~\bibnamefont {Daniel~Dapkus}},\ and\ \bibinfo {author} {\bibfnamefont {S.~R.}\ \bibnamefont {Nutt}},\ }\bibfield  {title} {\bibinfo {title} {Vertical nonpolar growth templates for light emitting diodes formed with gan nanosheets},\ }\href@noop {} {\bibfield  {journal} {\bibinfo  {journal} {Applied Physics Letters}\ }\textbf {\bibinfo {volume} {100}} (\bibinfo {year} {2012})}\BibitemShut {NoStop}%
\bibitem [{\citenamefont {Chen}\ \emph {et~al.}(2018)\citenamefont {Chen}, \citenamefont {Liu}, \citenamefont {Liu}, \citenamefont {Lv}, \citenamefont {Wei}, \citenamefont {Zhang}, \citenamefont {Zeng}, \citenamefont {Wang},\ and\ \citenamefont {Fu}}]{chen2018growth}%
  \BibitemOpen
  \bibfield  {author} {\bibinfo {author} {\bibfnamefont {Y.}~\bibnamefont {Chen}}, \bibinfo {author} {\bibfnamefont {K.}~\bibnamefont {Liu}}, \bibinfo {author} {\bibfnamefont {J.}~\bibnamefont {Liu}}, \bibinfo {author} {\bibfnamefont {T.}~\bibnamefont {Lv}}, \bibinfo {author} {\bibfnamefont {B.}~\bibnamefont {Wei}}, \bibinfo {author} {\bibfnamefont {T.}~\bibnamefont {Zhang}}, \bibinfo {author} {\bibfnamefont {M.}~\bibnamefont {Zeng}}, \bibinfo {author} {\bibfnamefont {Z.}~\bibnamefont {Wang}},\ and\ \bibinfo {author} {\bibfnamefont {L.}~\bibnamefont {Fu}},\ }\bibfield  {title} {\bibinfo {title} {Growth of 2d gan single crystals on liquid metals},\ }\href@noop {} {\bibfield  {journal} {\bibinfo  {journal} {Journal of the American Chemical Society}\ }\textbf {\bibinfo {volume} {140}},\ \bibinfo {pages} {16392} (\bibinfo {year} {2018})}\BibitemShut {NoStop}%
\bibitem [{\citenamefont {Chen}\ \emph {et~al.}(2019)\citenamefont {Chen}, \citenamefont {Liu}, \citenamefont {Liu}, \citenamefont {Si}, \citenamefont {Ding}, \citenamefont {Li}, \citenamefont {Lv}, \citenamefont {Liu},\ and\ \citenamefont {Fu}}]{chen2019gan}%
  \BibitemOpen
  \bibfield  {author} {\bibinfo {author} {\bibfnamefont {Y.}~\bibnamefont {Chen}}, \bibinfo {author} {\bibfnamefont {J.}~\bibnamefont {Liu}}, \bibinfo {author} {\bibfnamefont {K.}~\bibnamefont {Liu}}, \bibinfo {author} {\bibfnamefont {J.}~\bibnamefont {Si}}, \bibinfo {author} {\bibfnamefont {Y.}~\bibnamefont {Ding}}, \bibinfo {author} {\bibfnamefont {L.}~\bibnamefont {Li}}, \bibinfo {author} {\bibfnamefont {T.}~\bibnamefont {Lv}}, \bibinfo {author} {\bibfnamefont {J.}~\bibnamefont {Liu}},\ and\ \bibinfo {author} {\bibfnamefont {L.}~\bibnamefont {Fu}},\ }\bibfield  {title} {\bibinfo {title} {Gan in different dimensionalities: Properties, synthesis, and applications},\ }\href@noop {} {\bibfield  {journal} {\bibinfo  {journal} {Materials Science and Engineering: R: Reports}\ }\textbf {\bibinfo {volume} {138}},\ \bibinfo {pages} {60} (\bibinfo {year} {2019})}\BibitemShut {NoStop}%
\bibitem [{\citenamefont {Yue}\ \emph {et~al.}(2019)\citenamefont {Yue}, \citenamefont {Yang},\ and\ \citenamefont {Liao}}]{yue2019controlling}%
  \BibitemOpen
  \bibfield  {author} {\bibinfo {author} {\bibfnamefont {S.-Y.}\ \bibnamefont {Yue}}, \bibinfo {author} {\bibfnamefont {R.}~\bibnamefont {Yang}},\ and\ \bibinfo {author} {\bibfnamefont {B.}~\bibnamefont {Liao}},\ }\bibfield  {title} {\bibinfo {title} {Controlling thermal conductivity of two-dimensional materials via externally induced phonon-electron interaction},\ }\href@noop {} {\bibfield  {journal} {\bibinfo  {journal} {Physical Review B}\ }\textbf {\bibinfo {volume} {100}},\ \bibinfo {pages} {115408} (\bibinfo {year} {2019})}\BibitemShut {NoStop}%
\bibitem [{\citenamefont {Tong}\ \emph {et~al.}(2021)\citenamefont {Tong}, \citenamefont {Dumitrica},\ and\ \citenamefont {Frauenheim}}]{tong2021ultralow}%
  \BibitemOpen
  \bibfield  {author} {\bibinfo {author} {\bibfnamefont {Z.}~\bibnamefont {Tong}}, \bibinfo {author} {\bibfnamefont {T.}~\bibnamefont {Dumitrica}},\ and\ \bibinfo {author} {\bibfnamefont {T.}~\bibnamefont {Frauenheim}},\ }\bibfield  {title} {\bibinfo {title} {Ultralow thermal conductivity in two-dimensional moo3},\ }\href@noop {} {\bibfield  {journal} {\bibinfo  {journal} {Nano letters}\ }\textbf {\bibinfo {volume} {21}},\ \bibinfo {pages} {4351} (\bibinfo {year} {2021})}\BibitemShut {NoStop}%
\bibitem [{\citenamefont {Yang}\ \emph {et~al.}(2021)\citenamefont {Yang}, \citenamefont {Liu}, \citenamefont {Meng},\ and\ \citenamefont {Li}}]{yang2021tuning}%
  \BibitemOpen
  \bibfield  {author} {\bibinfo {author} {\bibfnamefont {X.}~\bibnamefont {Yang}}, \bibinfo {author} {\bibfnamefont {Z.}~\bibnamefont {Liu}}, \bibinfo {author} {\bibfnamefont {F.}~\bibnamefont {Meng}},\ and\ \bibinfo {author} {\bibfnamefont {W.}~\bibnamefont {Li}},\ }\bibfield  {title} {\bibinfo {title} {Tuning the phonon transport in bilayer graphene to an anomalous regime dominated by electron-phonon scattering},\ }\href@noop {} {\bibfield  {journal} {\bibinfo  {journal} {Physical Review B}\ }\textbf {\bibinfo {volume} {104}},\ \bibinfo {pages} {L100306} (\bibinfo {year} {2021})}\BibitemShut {NoStop}%
\bibitem [{\citenamefont {Liu}\ \emph {et~al.}(2020)\citenamefont {Liu}, \citenamefont {Yao}, \citenamefont {Yang}, \citenamefont {Xi},\ and\ \citenamefont {Ke}}]{liu2020strong}%
  \BibitemOpen
  \bibfield  {author} {\bibinfo {author} {\bibfnamefont {C.}~\bibnamefont {Liu}}, \bibinfo {author} {\bibfnamefont {M.}~\bibnamefont {Yao}}, \bibinfo {author} {\bibfnamefont {J.}~\bibnamefont {Yang}}, \bibinfo {author} {\bibfnamefont {J.}~\bibnamefont {Xi}},\ and\ \bibinfo {author} {\bibfnamefont {X.}~\bibnamefont {Ke}},\ }\bibfield  {title} {\bibinfo {title} {Strong electron-phonon interaction induced significant reduction in lattice thermal conductivities for single-layer mos2 and ptsse},\ }\href@noop {} {\bibfield  {journal} {\bibinfo  {journal} {Materials Today Physics}\ }\textbf {\bibinfo {volume} {15}},\ \bibinfo {pages} {100277} (\bibinfo {year} {2020})}\BibitemShut {NoStop}%
\bibitem [{\citenamefont {Guo}\ \emph {et~al.}(2024)\citenamefont {Guo}, \citenamefont {Yan}, \citenamefont {Sun}, \citenamefont {Pan}, \citenamefont {He}, \citenamefont {Zhang}, \citenamefont {Yang}, \citenamefont {Wang}, \citenamefont {Zhang}, \citenamefont {Li} \emph {et~al.}}]{guo2024four}%
  \BibitemOpen
  \bibfield  {author} {\bibinfo {author} {\bibfnamefont {H.}~\bibnamefont {Guo}}, \bibinfo {author} {\bibfnamefont {W.}~\bibnamefont {Yan}}, \bibinfo {author} {\bibfnamefont {J.}~\bibnamefont {Sun}}, \bibinfo {author} {\bibfnamefont {Y.}~\bibnamefont {Pan}}, \bibinfo {author} {\bibfnamefont {H.}~\bibnamefont {He}}, \bibinfo {author} {\bibfnamefont {Y.}~\bibnamefont {Zhang}}, \bibinfo {author} {\bibfnamefont {F.}~\bibnamefont {Yang}}, \bibinfo {author} {\bibfnamefont {Y.}~\bibnamefont {Wang}}, \bibinfo {author} {\bibfnamefont {C.}~\bibnamefont {Zhang}}, \bibinfo {author} {\bibfnamefont {R.}~\bibnamefont {Li}}, \emph {et~al.},\ }\bibfield  {title} {\bibinfo {title} {Four-phonon scattering and thermal transport in 2h--mote2},\ }\href@noop {} {\bibfield  {journal} {\bibinfo  {journal} {Materials Today Physics}\ }\textbf {\bibinfo {volume} {40}},\ \bibinfo {pages} {101314} (\bibinfo {year} {2024})}\BibitemShut {NoStop}%
\bibitem [{\citenamefont {Huang}\ \emph {et~al.}(2019)\citenamefont {Huang}, \citenamefont {Zhou}, \citenamefont {Wang},\ and\ \citenamefont {Sun}}]{huang2019abnormally}%
  \BibitemOpen
  \bibfield  {author} {\bibinfo {author} {\bibfnamefont {Y.}~\bibnamefont {Huang}}, \bibinfo {author} {\bibfnamefont {J.}~\bibnamefont {Zhou}}, \bibinfo {author} {\bibfnamefont {G.}~\bibnamefont {Wang}},\ and\ \bibinfo {author} {\bibfnamefont {Z.}~\bibnamefont {Sun}},\ }\bibfield  {title} {\bibinfo {title} {Abnormally strong electron-phonon scattering induced unprecedented reduction in lattice thermal conductivity of two-dimensional nb2c},\ }\href@noop {} {\bibfield  {journal} {\bibinfo  {journal} {Journal of the American Chemical Society}\ }\textbf {\bibinfo {volume} {141}},\ \bibinfo {pages} {8503} (\bibinfo {year} {2019})}\BibitemShut {NoStop}%
\bibitem [{\citenamefont {Jiang}\ \emph {et~al.}(2017)\citenamefont {Jiang}, \citenamefont {Cai}, \citenamefont {Tao}, \citenamefont {Wei}, \citenamefont {Bi},\ and\ \citenamefont {Chen}}]{jiang2017phonon}%
  \BibitemOpen
  \bibfield  {author} {\bibinfo {author} {\bibfnamefont {Y.}~\bibnamefont {Jiang}}, \bibinfo {author} {\bibfnamefont {S.}~\bibnamefont {Cai}}, \bibinfo {author} {\bibfnamefont {Y.}~\bibnamefont {Tao}}, \bibinfo {author} {\bibfnamefont {Z.}~\bibnamefont {Wei}}, \bibinfo {author} {\bibfnamefont {K.}~\bibnamefont {Bi}},\ and\ \bibinfo {author} {\bibfnamefont {Y.}~\bibnamefont {Chen}},\ }\bibfield  {title} {\bibinfo {title} {Phonon transport properties of bulk and monolayer gan from first-principles calculations},\ }\href@noop {} {\bibfield  {journal} {\bibinfo  {journal} {Computational Materials Science}\ }\textbf {\bibinfo {volume} {138}},\ \bibinfo {pages} {419} (\bibinfo {year} {2017})}\BibitemShut {NoStop}%
\bibitem [{\citenamefont {Qin}\ \emph {et~al.}(2017{\natexlab{a}})\citenamefont {Qin}, \citenamefont {Qin}, \citenamefont {Wang},\ and\ \citenamefont {Hu}}]{qin2017anomalously}%
  \BibitemOpen
  \bibfield  {author} {\bibinfo {author} {\bibfnamefont {G.}~\bibnamefont {Qin}}, \bibinfo {author} {\bibfnamefont {Z.}~\bibnamefont {Qin}}, \bibinfo {author} {\bibfnamefont {H.}~\bibnamefont {Wang}},\ and\ \bibinfo {author} {\bibfnamefont {M.}~\bibnamefont {Hu}},\ }\bibfield  {title} {\bibinfo {title} {Anomalously temperature-dependent thermal conductivity of monolayer gan with large deviations from the traditional 1/t law},\ }\href@noop {} {\bibfield  {journal} {\bibinfo  {journal} {Physical Review B}\ }\textbf {\bibinfo {volume} {95}},\ \bibinfo {pages} {195416} (\bibinfo {year} {2017}{\natexlab{a}})}\BibitemShut {NoStop}%
\bibitem [{\citenamefont {Qin}\ \emph {et~al.}(2017{\natexlab{b}})\citenamefont {Qin}, \citenamefont {Qin}, \citenamefont {Zuo}, \citenamefont {Xiong},\ and\ \citenamefont {Hu}}]{qin2017orbitally}%
  \BibitemOpen
  \bibfield  {author} {\bibinfo {author} {\bibfnamefont {Z.}~\bibnamefont {Qin}}, \bibinfo {author} {\bibfnamefont {G.}~\bibnamefont {Qin}}, \bibinfo {author} {\bibfnamefont {X.}~\bibnamefont {Zuo}}, \bibinfo {author} {\bibfnamefont {Z.}~\bibnamefont {Xiong}},\ and\ \bibinfo {author} {\bibfnamefont {M.}~\bibnamefont {Hu}},\ }\bibfield  {title} {\bibinfo {title} {Orbitally driven low thermal conductivity of monolayer gallium nitride (gan) with planar honeycomb structure: a comparative study},\ }\href@noop {} {\bibfield  {journal} {\bibinfo  {journal} {Nanoscale}\ }\textbf {\bibinfo {volume} {9}},\ \bibinfo {pages} {4295} (\bibinfo {year} {2017}{\natexlab{b}})}\BibitemShut {NoStop}%
\bibitem [{\citenamefont {Shen}\ \emph {et~al.}(2022)\citenamefont {Shen}, \citenamefont {Hadaeghi}, \citenamefont {Singh}, \citenamefont {Long}, \citenamefont {Fan}, \citenamefont {Qin},\ and\ \citenamefont {Zhang}}]{shen2022two}%
  \BibitemOpen
  \bibfield  {author} {\bibinfo {author} {\bibfnamefont {C.}~\bibnamefont {Shen}}, \bibinfo {author} {\bibfnamefont {N.}~\bibnamefont {Hadaeghi}}, \bibinfo {author} {\bibfnamefont {H.~K.}\ \bibnamefont {Singh}}, \bibinfo {author} {\bibfnamefont {T.}~\bibnamefont {Long}}, \bibinfo {author} {\bibfnamefont {L.}~\bibnamefont {Fan}}, \bibinfo {author} {\bibfnamefont {G.}~\bibnamefont {Qin}},\ and\ \bibinfo {author} {\bibfnamefont {H.}~\bibnamefont {Zhang}},\ }\bibfield  {title} {\bibinfo {title} {Two-dimensional buckling structure induces the ultra-low thermal conductivity: a comparative study of the group gax (x= n, p, as)},\ }\href@noop {} {\bibfield  {journal} {\bibinfo  {journal} {Journal of Materials Chemistry C}\ }\textbf {\bibinfo {volume} {10}},\ \bibinfo {pages} {1436} (\bibinfo {year} {2022})}\BibitemShut {NoStop}%
\bibitem [{\citenamefont {Cai}\ \emph {et~al.}(2021)\citenamefont {Cai}, \citenamefont {Sun}, \citenamefont {Xu}, \citenamefont {Ma},\ and\ \citenamefont {Xu}}]{cai2021effect}%
  \BibitemOpen
  \bibfield  {author} {\bibinfo {author} {\bibfnamefont {X.}~\bibnamefont {Cai}}, \bibinfo {author} {\bibfnamefont {G.}~\bibnamefont {Sun}}, \bibinfo {author} {\bibfnamefont {Y.}~\bibnamefont {Xu}}, \bibinfo {author} {\bibfnamefont {J.}~\bibnamefont {Ma}},\ and\ \bibinfo {author} {\bibfnamefont {D.}~\bibnamefont {Xu}},\ }\bibfield  {title} {\bibinfo {title} {Effect of hydrogenation on the thermal conductivity of 2d gallium nitride},\ }\href@noop {} {\bibfield  {journal} {\bibinfo  {journal} {Physical Chemistry Chemical Physics}\ }\textbf {\bibinfo {volume} {23}},\ \bibinfo {pages} {22423} (\bibinfo {year} {2021})}\BibitemShut {NoStop}%
\bibitem [{\citenamefont {Sun}\ \emph {et~al.}(2023{\natexlab{a}})\citenamefont {Sun}, \citenamefont {Ma}, \citenamefont {Liu}, \citenamefont {Xiang}, \citenamefont {Xu}, \citenamefont {Liu},\ and\ \citenamefont {Luo}}]{sun2023four}%
  \BibitemOpen
  \bibfield  {author} {\bibinfo {author} {\bibfnamefont {G.}~\bibnamefont {Sun}}, \bibinfo {author} {\bibfnamefont {J.}~\bibnamefont {Ma}}, \bibinfo {author} {\bibfnamefont {C.}~\bibnamefont {Liu}}, \bibinfo {author} {\bibfnamefont {Z.}~\bibnamefont {Xiang}}, \bibinfo {author} {\bibfnamefont {D.}~\bibnamefont {Xu}}, \bibinfo {author} {\bibfnamefont {T.-H.}\ \bibnamefont {Liu}},\ and\ \bibinfo {author} {\bibfnamefont {X.}~\bibnamefont {Luo}},\ }\bibfield  {title} {\bibinfo {title} {Four-phonon and normal scattering in 2d hexagonal structures},\ }\href@noop {} {\bibfield  {journal} {\bibinfo  {journal} {International Journal of Heat and Mass Transfer}\ }\textbf {\bibinfo {volume} {215}},\ \bibinfo {pages} {124475} (\bibinfo {year} {2023}{\natexlab{a}})}\BibitemShut {NoStop}%
\bibitem [{\citenamefont {Feng}\ and\ \citenamefont {Ruan}(2018)}]{feng2018four}%
  \BibitemOpen
  \bibfield  {author} {\bibinfo {author} {\bibfnamefont {T.}~\bibnamefont {Feng}}\ and\ \bibinfo {author} {\bibfnamefont {X.}~\bibnamefont {Ruan}},\ }\bibfield  {title} {\bibinfo {title} {Four-phonon scattering reduces intrinsic thermal conductivity of graphene and the contributions from flexural phonons},\ }\href@noop {} {\bibfield  {journal} {\bibinfo  {journal} {Physical Review B}\ }\textbf {\bibinfo {volume} {97}},\ \bibinfo {pages} {045202} (\bibinfo {year} {2018})}\BibitemShut {NoStop}%
\bibitem [{\citenamefont {Han}\ \emph {et~al.}(2022{\natexlab{a}})\citenamefont {Han}, \citenamefont {Yang}, \citenamefont {Sullivan}, \citenamefont {Feng}, \citenamefont {Shi}, \citenamefont {Li},\ and\ \citenamefont {Ruan}}]{han2022raman}%
  \BibitemOpen
  \bibfield  {author} {\bibinfo {author} {\bibfnamefont {Z.}~\bibnamefont {Han}}, \bibinfo {author} {\bibfnamefont {X.}~\bibnamefont {Yang}}, \bibinfo {author} {\bibfnamefont {S.~E.}\ \bibnamefont {Sullivan}}, \bibinfo {author} {\bibfnamefont {T.}~\bibnamefont {Feng}}, \bibinfo {author} {\bibfnamefont {L.}~\bibnamefont {Shi}}, \bibinfo {author} {\bibfnamefont {W.}~\bibnamefont {Li}},\ and\ \bibinfo {author} {\bibfnamefont {X.}~\bibnamefont {Ruan}},\ }\bibfield  {title} {\bibinfo {title} {Raman linewidth contributions from four-phonon and electron-phonon interactions in graphene},\ }\href@noop {} {\bibfield  {journal} {\bibinfo  {journal} {Physical Review Letters}\ }\textbf {\bibinfo {volume} {128}},\ \bibinfo {pages} {045901} (\bibinfo {year} {2022}{\natexlab{a}})}\BibitemShut {NoStop}%
\bibitem [{\citenamefont {Li}\ \emph {et~al.}(2024)\citenamefont {Li}, \citenamefont {Tang}, \citenamefont {Zheng}, \citenamefont {Wang}, \citenamefont {Cui}, \citenamefont {Xu}, \citenamefont {Xu}, \citenamefont {Liu}, \citenamefont {Zhu}, \citenamefont {Guo} \emph {et~al.}}]{li2024convergent}%
  \BibitemOpen
  \bibfield  {author} {\bibinfo {author} {\bibfnamefont {G.}~\bibnamefont {Li}}, \bibinfo {author} {\bibfnamefont {J.}~\bibnamefont {Tang}}, \bibinfo {author} {\bibfnamefont {J.}~\bibnamefont {Zheng}}, \bibinfo {author} {\bibfnamefont {Q.}~\bibnamefont {Wang}}, \bibinfo {author} {\bibfnamefont {Z.}~\bibnamefont {Cui}}, \bibinfo {author} {\bibfnamefont {K.}~\bibnamefont {Xu}}, \bibinfo {author} {\bibfnamefont {J.}~\bibnamefont {Xu}}, \bibinfo {author} {\bibfnamefont {T.-H.}\ \bibnamefont {Liu}}, \bibinfo {author} {\bibfnamefont {G.}~\bibnamefont {Zhu}}, \bibinfo {author} {\bibfnamefont {R.}~\bibnamefont {Guo}}, \emph {et~al.},\ }\bibfield  {title} {\bibinfo {title} {Convergent thermal conductivity in strained monolayer graphene},\ }\href@noop {} {\bibfield  {journal} {\bibinfo  {journal} {Physical Review B}\ }\textbf {\bibinfo {volume} {109}},\ \bibinfo {pages} {035420} (\bibinfo {year} {2024})}\BibitemShut {NoStop}%
\bibitem [{\citenamefont {Karaaslan}\ \emph {et~al.}(2020)\citenamefont {Karaaslan}, \citenamefont {Yapicioglu},\ and\ \citenamefont {Sevik}}]{karaaslan2020assessment}%
  \BibitemOpen
  \bibfield  {author} {\bibinfo {author} {\bibfnamefont {Y.}~\bibnamefont {Karaaslan}}, \bibinfo {author} {\bibfnamefont {H.}~\bibnamefont {Yapicioglu}},\ and\ \bibinfo {author} {\bibfnamefont {C.}~\bibnamefont {Sevik}},\ }\bibfield  {title} {\bibinfo {title} {Assessment of thermal transport properties of group-iii nitrides: A classical molecular dynamics study with transferable tersoff-type interatomic potentials},\ }\href@noop {} {\bibfield  {journal} {\bibinfo  {journal} {Physical Review Applied}\ }\textbf {\bibinfo {volume} {13}},\ \bibinfo {pages} {034027} (\bibinfo {year} {2020})}\BibitemShut {NoStop}%
\bibitem [{\citenamefont {Zhou}\ \emph {et~al.}(2023)\citenamefont {Zhou}, \citenamefont {Zhou}, \citenamefont {Hua}, \citenamefont {Bawane},\ and\ \citenamefont {Feng}}]{zhou2023extreme}%
  \BibitemOpen
  \bibfield  {author} {\bibinfo {author} {\bibfnamefont {H.}~\bibnamefont {Zhou}}, \bibinfo {author} {\bibfnamefont {S.}~\bibnamefont {Zhou}}, \bibinfo {author} {\bibfnamefont {Z.}~\bibnamefont {Hua}}, \bibinfo {author} {\bibfnamefont {K.}~\bibnamefont {Bawane}},\ and\ \bibinfo {author} {\bibfnamefont {T.}~\bibnamefont {Feng}},\ }\bibfield  {title} {\bibinfo {title} {Extreme sensitivity of higher-order interatomic force constants and thermal conductivity to the energy surface roughness of exchange-correlation functionals},\ }\href@noop {} {\bibfield  {journal} {\bibinfo  {journal} {Applied Physics Letters}\ }\textbf {\bibinfo {volume} {123}} (\bibinfo {year} {2023})}\BibitemShut {NoStop}%
\bibitem [{\citenamefont {Dongre}\ \emph {et~al.}(2022)\citenamefont {Dongre}, \citenamefont {Carrete}, \citenamefont {Mingo},\ and\ \citenamefont {Madsen}}]{dongre2022thermal}%
  \BibitemOpen
  \bibfield  {author} {\bibinfo {author} {\bibfnamefont {B.}~\bibnamefont {Dongre}}, \bibinfo {author} {\bibfnamefont {J.}~\bibnamefont {Carrete}}, \bibinfo {author} {\bibfnamefont {N.}~\bibnamefont {Mingo}},\ and\ \bibinfo {author} {\bibfnamefont {G.~K.}\ \bibnamefont {Madsen}},\ }\bibfield  {title} {\bibinfo {title} {Thermal conductivity of group-iii phosphides: The special case of gap},\ }\href@noop {} {\bibfield  {journal} {\bibinfo  {journal} {Physical Review B}\ }\textbf {\bibinfo {volume} {106}},\ \bibinfo {pages} {205202} (\bibinfo {year} {2022})}\BibitemShut {NoStop}%
\bibitem [{\citenamefont {Li}\ \emph {et~al.}(2023)\citenamefont {Li}, \citenamefont {Xia},\ and\ \citenamefont {Wolverton}}]{li2023first}%
  \BibitemOpen
  \bibfield  {author} {\bibinfo {author} {\bibfnamefont {Z.}~\bibnamefont {Li}}, \bibinfo {author} {\bibfnamefont {Y.}~\bibnamefont {Xia}},\ and\ \bibinfo {author} {\bibfnamefont {C.}~\bibnamefont {Wolverton}},\ }\bibfield  {title} {\bibinfo {title} {First-principles calculations of lattice thermal conductivity in tl 3 vse 4: Uncertainties from different approaches of force constants},\ }\href@noop {} {\bibfield  {journal} {\bibinfo  {journal} {Physical Review B}\ }\textbf {\bibinfo {volume} {108}},\ \bibinfo {pages} {184307} (\bibinfo {year} {2023})}\BibitemShut {NoStop}%
\bibitem [{\citenamefont {Ouyang}\ \emph {et~al.}(2023)\citenamefont {Ouyang}, \citenamefont {Zeng}, \citenamefont {Wang}, \citenamefont {Wang},\ and\ \citenamefont {Chen}}]{ouyang2023role}%
  \BibitemOpen
  \bibfield  {author} {\bibinfo {author} {\bibfnamefont {N.}~\bibnamefont {Ouyang}}, \bibinfo {author} {\bibfnamefont {Z.}~\bibnamefont {Zeng}}, \bibinfo {author} {\bibfnamefont {C.}~\bibnamefont {Wang}}, \bibinfo {author} {\bibfnamefont {Q.}~\bibnamefont {Wang}},\ and\ \bibinfo {author} {\bibfnamefont {Y.}~\bibnamefont {Chen}},\ }\bibfield  {title} {\bibinfo {title} {Role of high-order lattice anharmonicity in the phonon thermal transport of silver halide ag x (x= cl, br, i)},\ }\href@noop {} {\bibfield  {journal} {\bibinfo  {journal} {Physical Review B}\ }\textbf {\bibinfo {volume} {108}},\ \bibinfo {pages} {174302} (\bibinfo {year} {2023})}\BibitemShut {NoStop}%
\bibitem [{\citenamefont {Wang}\ \emph {et~al.}(2022)\citenamefont {Wang}, \citenamefont {Li}, \citenamefont {Zhang},\ and\ \citenamefont {Bao}}]{wang2022roles}%
  \BibitemOpen
  \bibfield  {author} {\bibinfo {author} {\bibfnamefont {A.}~\bibnamefont {Wang}}, \bibinfo {author} {\bibfnamefont {S.}~\bibnamefont {Li}}, \bibinfo {author} {\bibfnamefont {X.}~\bibnamefont {Zhang}},\ and\ \bibinfo {author} {\bibfnamefont {H.}~\bibnamefont {Bao}},\ }\bibfield  {title} {\bibinfo {title} {Roles of electrons on the thermal transport of 2d metallic mxenes},\ }\href@noop {} {\bibfield  {journal} {\bibinfo  {journal} {Physical Review Materials}\ }\textbf {\bibinfo {volume} {6}},\ \bibinfo {pages} {014009} (\bibinfo {year} {2022})}\BibitemShut {NoStop}%
\bibitem [{\citenamefont {Giannozzi}\ \emph {et~al.}(2009)\citenamefont {Giannozzi}, \citenamefont {Baroni}, \citenamefont {Bonini}, \citenamefont {Calandra}, \citenamefont {Car}, \citenamefont {Cavazzoni}, \citenamefont {Ceresoli}, \citenamefont {Chiarotti}, \citenamefont {Cococcioni}, \citenamefont {Dabo} \emph {et~al.}}]{giannozzi2009quantum}%
  \BibitemOpen
  \bibfield  {author} {\bibinfo {author} {\bibfnamefont {P.}~\bibnamefont {Giannozzi}}, \bibinfo {author} {\bibfnamefont {S.}~\bibnamefont {Baroni}}, \bibinfo {author} {\bibfnamefont {N.}~\bibnamefont {Bonini}}, \bibinfo {author} {\bibfnamefont {M.}~\bibnamefont {Calandra}}, \bibinfo {author} {\bibfnamefont {R.}~\bibnamefont {Car}}, \bibinfo {author} {\bibfnamefont {C.}~\bibnamefont {Cavazzoni}}, \bibinfo {author} {\bibfnamefont {D.}~\bibnamefont {Ceresoli}}, \bibinfo {author} {\bibfnamefont {G.~L.}\ \bibnamefont {Chiarotti}}, \bibinfo {author} {\bibfnamefont {M.}~\bibnamefont {Cococcioni}}, \bibinfo {author} {\bibfnamefont {I.}~\bibnamefont {Dabo}}, \emph {et~al.},\ }\bibfield  {title} {\bibinfo {title} {Quantum espresso: a modular and open-source software project for quantum simulations of materials},\ }\href@noop {} {\bibfield  {journal} {\bibinfo  {journal} {Journal of physics: Condensed matter}\ }\textbf {\bibinfo {volume} {21}},\ \bibinfo {pages} {395502} (\bibinfo {year} {2009})}\BibitemShut {NoStop}%
\bibitem [{\citenamefont {Jackson}\ and\ \citenamefont {Pederson}(1990)}]{jackson1990accurate}%
  \BibitemOpen
  \bibfield  {author} {\bibinfo {author} {\bibfnamefont {K.}~\bibnamefont {Jackson}}\ and\ \bibinfo {author} {\bibfnamefont {M.~R.}\ \bibnamefont {Pederson}},\ }\bibfield  {title} {\bibinfo {title} {Accurate forces in a local-orbital approach to the local-density approximation},\ }\href@noop {} {\bibfield  {journal} {\bibinfo  {journal} {Physical Review B}\ }\textbf {\bibinfo {volume} {42}},\ \bibinfo {pages} {3276} (\bibinfo {year} {1990})}\BibitemShut {NoStop}%
\bibitem [{\citenamefont {Hamann}(2013)}]{hamann2013optimized}%
  \BibitemOpen
  \bibfield  {author} {\bibinfo {author} {\bibfnamefont {D.}~\bibnamefont {Hamann}},\ }\bibfield  {title} {\bibinfo {title} {Optimized norm-conserving vanderbilt pseudopotentials},\ }\href@noop {} {\bibfield  {journal} {\bibinfo  {journal} {Physical Review B}\ }\textbf {\bibinfo {volume} {88}},\ \bibinfo {pages} {085117} (\bibinfo {year} {2013})}\BibitemShut {NoStop}%
\bibitem [{\citenamefont {van Setten}\ \emph {et~al.}(2018)\citenamefont {van Setten}, \citenamefont {Giantomassi}, \citenamefont {Bousquet}, \citenamefont {Verstraete}, \citenamefont {Hamann}, \citenamefont {Gonze},\ and\ \citenamefont {Rignanese}}]{van2018pseudodojo}%
  \BibitemOpen
  \bibfield  {author} {\bibinfo {author} {\bibfnamefont {M.~J.}\ \bibnamefont {van Setten}}, \bibinfo {author} {\bibfnamefont {M.}~\bibnamefont {Giantomassi}}, \bibinfo {author} {\bibfnamefont {E.}~\bibnamefont {Bousquet}}, \bibinfo {author} {\bibfnamefont {M.~J.}\ \bibnamefont {Verstraete}}, \bibinfo {author} {\bibfnamefont {D.~R.}\ \bibnamefont {Hamann}}, \bibinfo {author} {\bibfnamefont {X.}~\bibnamefont {Gonze}},\ and\ \bibinfo {author} {\bibfnamefont {G.-M.}\ \bibnamefont {Rignanese}},\ }\bibfield  {title} {\bibinfo {title} {The pseudodojo: Training and grading a 85 element optimized norm-conserving pseudopotential table},\ }\href@noop {} {\bibfield  {journal} {\bibinfo  {journal} {Computer Physics Communications}\ }\textbf {\bibinfo {volume} {226}},\ \bibinfo {pages} {39} (\bibinfo {year} {2018})}\BibitemShut {NoStop}%
\bibitem [{\citenamefont {Fletcher}(1970)}]{fletcher1970new}%
  \BibitemOpen
  \bibfield  {author} {\bibinfo {author} {\bibfnamefont {R.}~\bibnamefont {Fletcher}},\ }\bibfield  {title} {\bibinfo {title} {A new approach to variable metric algorithms},\ }\href@noop {} {\bibfield  {journal} {\bibinfo  {journal} {The computer journal}\ }\textbf {\bibinfo {volume} {13}},\ \bibinfo {pages} {317} (\bibinfo {year} {1970})}\BibitemShut {NoStop}%
\bibitem [{\citenamefont {Broyden}(1970)}]{broyden1970convergence}%
  \BibitemOpen
  \bibfield  {author} {\bibinfo {author} {\bibfnamefont {C.~G.}\ \bibnamefont {Broyden}},\ }\bibfield  {title} {\bibinfo {title} {The convergence of a class of double-rank minimization algorithms 1. general considerations},\ }\href@noop {} {\bibfield  {journal} {\bibinfo  {journal} {IMA Journal of Applied Mathematics}\ }\textbf {\bibinfo {volume} {6}},\ \bibinfo {pages} {76} (\bibinfo {year} {1970})}\BibitemShut {NoStop}%
\bibitem [{\citenamefont {Goldfarb}(1970)}]{goldfarb1970family}%
  \BibitemOpen
  \bibfield  {author} {\bibinfo {author} {\bibfnamefont {D.}~\bibnamefont {Goldfarb}},\ }\bibfield  {title} {\bibinfo {title} {A family of variable-metric methods derived by variational means},\ }\href@noop {} {\bibfield  {journal} {\bibinfo  {journal} {Mathematics of computation}\ }\textbf {\bibinfo {volume} {24}},\ \bibinfo {pages} {23} (\bibinfo {year} {1970})}\BibitemShut {NoStop}%
\bibitem [{\citenamefont {Shanno}(1970)}]{shanno1970conditioning}%
  \BibitemOpen
  \bibfield  {author} {\bibinfo {author} {\bibfnamefont {D.~F.}\ \bibnamefont {Shanno}},\ }\bibfield  {title} {\bibinfo {title} {Conditioning of quasi-newton methods for function minimization},\ }\href@noop {} {\bibfield  {journal} {\bibinfo  {journal} {Mathematics of computation}\ }\textbf {\bibinfo {volume} {24}},\ \bibinfo {pages} {647} (\bibinfo {year} {1970})}\BibitemShut {NoStop}%
\bibitem [{\citenamefont {Al~Balushi}\ \emph {et~al.}(2016)\citenamefont {Al~Balushi}, \citenamefont {Wang}, \citenamefont {Ghosh}, \citenamefont {Vil{\'a}}, \citenamefont {Eichfeld}, \citenamefont {Caldwell}, \citenamefont {Qin}, \citenamefont {Lin}, \citenamefont {DeSario}, \citenamefont {Stone} \emph {et~al.}}]{al2016two}%
  \BibitemOpen
  \bibfield  {author} {\bibinfo {author} {\bibfnamefont {Z.~Y.}\ \bibnamefont {Al~Balushi}}, \bibinfo {author} {\bibfnamefont {K.}~\bibnamefont {Wang}}, \bibinfo {author} {\bibfnamefont {R.~K.}\ \bibnamefont {Ghosh}}, \bibinfo {author} {\bibfnamefont {R.~A.}\ \bibnamefont {Vil{\'a}}}, \bibinfo {author} {\bibfnamefont {S.~M.}\ \bibnamefont {Eichfeld}}, \bibinfo {author} {\bibfnamefont {J.~D.}\ \bibnamefont {Caldwell}}, \bibinfo {author} {\bibfnamefont {X.}~\bibnamefont {Qin}}, \bibinfo {author} {\bibfnamefont {Y.-C.}\ \bibnamefont {Lin}}, \bibinfo {author} {\bibfnamefont {P.~A.}\ \bibnamefont {DeSario}}, \bibinfo {author} {\bibfnamefont {G.}~\bibnamefont {Stone}}, \emph {et~al.},\ }\bibfield  {title} {\bibinfo {title} {Two-dimensional gallium nitride realized via graphene encapsulation},\ }\href@noop {} {\bibfield  {journal} {\bibinfo  {journal} {Nature materials}\ }\textbf {\bibinfo {volume} {15}},\ \bibinfo {pages} {1166} (\bibinfo {year} {2016})}\BibitemShut {NoStop}%
\bibitem [{\citenamefont {Fugallo}\ \emph {et~al.}(2013)\citenamefont {Fugallo}, \citenamefont {Lazzeri}, \citenamefont {Paulatto},\ and\ \citenamefont {Mauri}}]{fugallo2013ab}%
  \BibitemOpen
  \bibfield  {author} {\bibinfo {author} {\bibfnamefont {G.}~\bibnamefont {Fugallo}}, \bibinfo {author} {\bibfnamefont {M.}~\bibnamefont {Lazzeri}}, \bibinfo {author} {\bibfnamefont {L.}~\bibnamefont {Paulatto}},\ and\ \bibinfo {author} {\bibfnamefont {F.}~\bibnamefont {Mauri}},\ }\bibfield  {title} {\bibinfo {title} {Ab initio variational approach for evaluating lattice thermal conductivity},\ }\href@noop {} {\bibfield  {journal} {\bibinfo  {journal} {Physical Review B}\ }\textbf {\bibinfo {volume} {88}},\ \bibinfo {pages} {045430} (\bibinfo {year} {2013})}\BibitemShut {NoStop}%
\bibitem [{\citenamefont {Baroni}\ \emph {et~al.}(2001)\citenamefont {Baroni}, \citenamefont {De~Gironcoli}, \citenamefont {Dal~Corso},\ and\ \citenamefont {Giannozzi}}]{baroni2001phonons}%
  \BibitemOpen
  \bibfield  {author} {\bibinfo {author} {\bibfnamefont {S.}~\bibnamefont {Baroni}}, \bibinfo {author} {\bibfnamefont {S.}~\bibnamefont {De~Gironcoli}}, \bibinfo {author} {\bibfnamefont {A.}~\bibnamefont {Dal~Corso}},\ and\ \bibinfo {author} {\bibfnamefont {P.}~\bibnamefont {Giannozzi}},\ }\bibfield  {title} {\bibinfo {title} {Phonons and related crystal properties from density-functional perturbation theory},\ }\href@noop {} {\bibfield  {journal} {\bibinfo  {journal} {Reviews of modern Physics}\ }\textbf {\bibinfo {volume} {73}},\ \bibinfo {pages} {515} (\bibinfo {year} {2001})}\BibitemShut {NoStop}%
\bibitem [{\citenamefont {Li}\ \emph {et~al.}(2014)\citenamefont {Li}, \citenamefont {Carrete}, \citenamefont {Katcho},\ and\ \citenamefont {Mingo}}]{li2014shengbte}%
  \BibitemOpen
  \bibfield  {author} {\bibinfo {author} {\bibfnamefont {W.}~\bibnamefont {Li}}, \bibinfo {author} {\bibfnamefont {J.}~\bibnamefont {Carrete}}, \bibinfo {author} {\bibfnamefont {N.~A.}\ \bibnamefont {Katcho}},\ and\ \bibinfo {author} {\bibfnamefont {N.}~\bibnamefont {Mingo}},\ }\bibfield  {title} {\bibinfo {title} {Shengbte: A solver of the boltzmann transport equation for phonons},\ }\href@noop {} {\bibfield  {journal} {\bibinfo  {journal} {Computer Physics Communications}\ }\textbf {\bibinfo {volume} {185}},\ \bibinfo {pages} {1747} (\bibinfo {year} {2014})}\BibitemShut {NoStop}%
\bibitem [{\citenamefont {Han}\ \emph {et~al.}(2022{\natexlab{b}})\citenamefont {Han}, \citenamefont {Yang}, \citenamefont {Li}, \citenamefont {Feng},\ and\ \citenamefont {Ruan}}]{han2022fourphonon}%
  \BibitemOpen
  \bibfield  {author} {\bibinfo {author} {\bibfnamefont {Z.}~\bibnamefont {Han}}, \bibinfo {author} {\bibfnamefont {X.}~\bibnamefont {Yang}}, \bibinfo {author} {\bibfnamefont {W.}~\bibnamefont {Li}}, \bibinfo {author} {\bibfnamefont {T.}~\bibnamefont {Feng}},\ and\ \bibinfo {author} {\bibfnamefont {X.}~\bibnamefont {Ruan}},\ }\bibfield  {title} {\bibinfo {title} {Fourphonon: An extension module to shengbte for computing four-phonon scattering rates and thermal conductivity},\ }\href@noop {} {\bibfield  {journal} {\bibinfo  {journal} {Computer Physics Communications}\ }\textbf {\bibinfo {volume} {270}},\ \bibinfo {pages} {108179} (\bibinfo {year} {2022}{\natexlab{b}})}\BibitemShut {NoStop}%
\bibitem [{\citenamefont {Marzari}\ \emph {et~al.}(2012)\citenamefont {Marzari}, \citenamefont {Mostofi}, \citenamefont {Yates}, \citenamefont {Souza},\ and\ \citenamefont {Vanderbilt}}]{marzari2012maximally}%
  \BibitemOpen
  \bibfield  {author} {\bibinfo {author} {\bibfnamefont {N.}~\bibnamefont {Marzari}}, \bibinfo {author} {\bibfnamefont {A.~A.}\ \bibnamefont {Mostofi}}, \bibinfo {author} {\bibfnamefont {J.~R.}\ \bibnamefont {Yates}}, \bibinfo {author} {\bibfnamefont {I.}~\bibnamefont {Souza}},\ and\ \bibinfo {author} {\bibfnamefont {D.}~\bibnamefont {Vanderbilt}},\ }\bibfield  {title} {\bibinfo {title} {Maximally localized wannier functions: Theory and applications},\ }\href@noop {} {\bibfield  {journal} {\bibinfo  {journal} {Reviews of Modern Physics}\ }\textbf {\bibinfo {volume} {84}},\ \bibinfo {pages} {1419} (\bibinfo {year} {2012})}\BibitemShut {NoStop}%
\bibitem [{\citenamefont {Tamura}(1983)}]{tamura1983isotope}%
  \BibitemOpen
  \bibfield  {author} {\bibinfo {author} {\bibfnamefont {S.-i.}\ \bibnamefont {Tamura}},\ }\bibfield  {title} {\bibinfo {title} {Isotope scattering of dispersive phonons in ge},\ }\href@noop {} {\bibfield  {journal} {\bibinfo  {journal} {Physical Review B}\ }\textbf {\bibinfo {volume} {27}},\ \bibinfo {pages} {858} (\bibinfo {year} {1983})}\BibitemShut {NoStop}%
\bibitem [{\citenamefont {Sun}\ \emph {et~al.}(2023{\natexlab{b}})\citenamefont {Sun}, \citenamefont {Chen}, \citenamefont {Li},\ and\ \citenamefont {Liu}}]{sun2023light}%
  \BibitemOpen
  \bibfield  {author} {\bibinfo {author} {\bibfnamefont {J.}~\bibnamefont {Sun}}, \bibinfo {author} {\bibfnamefont {G.}~\bibnamefont {Chen}}, \bibinfo {author} {\bibfnamefont {S.}~\bibnamefont {Li}},\ and\ \bibinfo {author} {\bibfnamefont {X.}~\bibnamefont {Liu}},\ }\bibfield  {title} {\bibinfo {title} {Light atomic mass induces low lattice thermal conductivity in janus transition-metal dichalcogenides msse (m═ mo, w)},\ }\href@noop {} {\bibfield  {journal} {\bibinfo  {journal} {The Journal of Physical Chemistry C}\ }\textbf {\bibinfo {volume} {127}},\ \bibinfo {pages} {17567} (\bibinfo {year} {2023}{\natexlab{b}})}\BibitemShut {NoStop}%
\bibitem [{\citenamefont {Li}\ \emph {et~al.}(2019)\citenamefont {Li}, \citenamefont {Tong},\ and\ \citenamefont {Bao}}]{li2019resolving}%
  \BibitemOpen
  \bibfield  {author} {\bibinfo {author} {\bibfnamefont {S.}~\bibnamefont {Li}}, \bibinfo {author} {\bibfnamefont {Z.}~\bibnamefont {Tong}},\ and\ \bibinfo {author} {\bibfnamefont {H.}~\bibnamefont {Bao}},\ }\bibfield  {title} {\bibinfo {title} {Resolving different scattering effects on the thermal and electrical transport in doped snse},\ }\href@noop {} {\bibfield  {journal} {\bibinfo  {journal} {Journal of Applied Physics}\ }\textbf {\bibinfo {volume} {126}} (\bibinfo {year} {2019})}\BibitemShut {NoStop}%
\bibitem [{\citenamefont {Kocaba{\c{s}}}\ \emph {et~al.}(2023)\citenamefont {Kocaba{\c{s}}}, \citenamefont {Ke{\c{c}}eli}, \citenamefont {V{\'a}zquez-Mayagoitia},\ and\ \citenamefont {Sevik}}]{kocabacs2023gaussian}%
  \BibitemOpen
  \bibfield  {author} {\bibinfo {author} {\bibfnamefont {T.}~\bibnamefont {Kocaba{\c{s}}}}, \bibinfo {author} {\bibfnamefont {M.}~\bibnamefont {Ke{\c{c}}eli}}, \bibinfo {author} {\bibfnamefont {{\'A}.}~\bibnamefont {V{\'a}zquez-Mayagoitia}},\ and\ \bibinfo {author} {\bibfnamefont {C.}~\bibnamefont {Sevik}},\ }\bibfield  {title} {\bibinfo {title} {Gaussian approximation potentials for accurate thermal properties of two-dimensional materials},\ }\href@noop {} {\bibfield  {journal} {\bibinfo  {journal} {Nanoscale}\ }\textbf {\bibinfo {volume} {15}},\ \bibinfo {pages} {8772} (\bibinfo {year} {2023})}\BibitemShut {NoStop}%
\bibitem [{\citenamefont {Han}\ and\ \citenamefont {Ruan}(2023)}]{han2023thermal}%
  \BibitemOpen
  \bibfield  {author} {\bibinfo {author} {\bibfnamefont {Z.}~\bibnamefont {Han}}\ and\ \bibinfo {author} {\bibfnamefont {X.}~\bibnamefont {Ruan}},\ }\bibfield  {title} {\bibinfo {title} {Thermal conductivity of monolayer graphene: Convergent and lower than diamond},\ }\href@noop {} {\bibfield  {journal} {\bibinfo  {journal} {Physical Review B}\ }\textbf {\bibinfo {volume} {108}},\ \bibinfo {pages} {L121412} (\bibinfo {year} {2023})}\BibitemShut {NoStop}%
\bibitem [{\citenamefont {Sun}\ \emph {et~al.}(2024)\citenamefont {Sun}, \citenamefont {Li}, \citenamefont {Tong}, \citenamefont {Shao}, \citenamefont {Chen}, \citenamefont {Liu}, \citenamefont {Xiong}, \citenamefont {An},\ and\ \citenamefont {Liu}}]{sun2024unexpected}%
  \BibitemOpen
  \bibfield  {author} {\bibinfo {author} {\bibfnamefont {J.}~\bibnamefont {Sun}}, \bibinfo {author} {\bibfnamefont {S.}~\bibnamefont {Li}}, \bibinfo {author} {\bibfnamefont {Z.}~\bibnamefont {Tong}}, \bibinfo {author} {\bibfnamefont {C.}~\bibnamefont {Shao}}, \bibinfo {author} {\bibfnamefont {X.}~\bibnamefont {Chen}}, \bibinfo {author} {\bibfnamefont {Q.}~\bibnamefont {Liu}}, \bibinfo {author} {\bibfnamefont {Y.}~\bibnamefont {Xiong}}, \bibinfo {author} {\bibfnamefont {M.}~\bibnamefont {An}},\ and\ \bibinfo {author} {\bibfnamefont {X.}~\bibnamefont {Liu}},\ }\bibfield  {title} {\bibinfo {title} {Unexpected weak effects of phonon-electron interactions on the lattice thermal conductivity of high-concentration n-type gan},\ }\href@noop {} {\bibfield  {journal} {\bibinfo  {journal} {arXiv preprint arXiv:2401.02133}\ } (\bibinfo {year} {2024})}\BibitemShut {NoStop}%
\end{thebibliography}%
\end{document}


\preprint{APS/123-QED}
\renewcommand{\thetable}{S\arabic{table}}
\renewcommand{\thefigure}{S\arabic{figure}}
\renewcommand{\theequation}{S\arabic{equation}}
\title{Revisiting phonon thermal transport in two-dimensional gallium nitride: higher-order phonon-phonon and phonon-electron scattering\\[6pt]
Supplemental Material}


\author{Jianshi Sun}
\affiliation{Institute of Micro/Nano Electromechanical System, College of Mechanical Engineering, Donghua University, Shanghai 201620, China
}%

\author{Xiangjun Liu}
\affiliation{%
 Institute of Micro/Nano Electromechanical System, College of Mechanical Engineering, Donghua University, Shanghai 201620, China
}%

\author{Yucheng Xiong}
\affiliation{%
Institute of Micro/Nano Electromechanical System, College of Mechanical Engineering, Donghua University, Shanghai 201620, China
}

\author{Yuhang Yao}
\affiliation{%
Institute of Micro/Nano Electromechanical System, College of Mechanical Engineering, Donghua University, Shanghai 201620, China
}

\author{Xiaolong Yang}
\affiliation{%
 College of Physics and Center of Quantum Materials \& Devices, Chongqing University, Chongqing 401331, China
}%

\author{Cheng Shao}
\affiliation{
 Thermal Science Research Center, Shandong Institute of Advanced Technology, Jinan, Shandong 250103, China
}

\author{Shouhang Li}
\email{shouhang.li@dhu.edu.cn}
\affiliation{Institute of Micro/Nano Electromechanical System, College of Mechanical Engineering, Donghua University, Shanghai 201620, China
}%


\date{\today}
\maketitle

\section{Computational Details}
Combining the linearized phonon Boltzmann transport equation and Fourier's law, the lattice thermal conductivity can be expressed as,\cite{broido2007intrinsic}
\begin{equation}
    \kappa_{\text {lat}, \alpha \beta}=\sum_{\lambda} c_{v, \lambda} v_{\lambda, \alpha} v_{\lambda, \beta} \tau_{\lambda}=\frac{1}{V} \sum_{\lambda} \hbar \omega_{\lambda} \frac{\partial n_{\lambda}}{\partial T} v_{\lambda, \alpha} v_{\lambda, \beta} \tau_{\lambda},
    \label{SE1}
\end{equation}
where $\alpha$ and $\beta$ are the Cartesian coordinates, $\lambda \equiv(\mathbf{q}, v)$ denotes the phonon mode with wave vector $\mathbf{q}$ and phonon polarization $v$, $c_{v, \lambda}$ is the phonon specific heat capacity, $v_{\lambda}$ is the phonon group velocity, and $\tau_{\lambda}$ is the phonon relaxation time. ${V}$ is the volume of the primitive cell, $\hbar$ is the reduced Planck’s constant, $\omega_{\lambda}$ is the phonon frequency, and $n_{\lambda}$ is the Bose-Einstein distribution at temperature ${T}$.

The essential step is to compute $\tau_{\lambda}$, which is associated with various phonon scattering processes. Using Matthiessen’s rules, the effective scattering rates can be obtained by considering three-phonon, four-phonon, phonon-isotope, and phonon-electron scattering:
\begin{equation}
    \frac{1}{\tau_{\lambda}}=\frac{1}{\tau_{\lambda}^{3 \mathrm{ph}}}+\frac{1}{\tau_{\lambda}^{4 \mathrm{ph}}}+\frac{1}{\tau_{\lambda}^{\mathrm{ph}-\mathrm{iso}}}+\frac{1}{\tau_{\lambda}^{\mathrm{ph}-\mathrm{el}}},
    \label{SE2}
\end{equation}

According to Fermi's golden rule\cite{albers1976normal}, the scattering rates of three-phonon and four-phonon can be expressed as,\cite{feng2016quantum}
\begin{equation}
    \frac{1}{\tau_{\lambda}^{\mathrm{3ph}}} = 2 \pi \sum_{\lambda_{1} \lambda_{2}}\left|V_{\lambda \lambda_{1} \lambda_{2}}\right|^{2}
    \left[
    \frac{1}{2}\left(1+n_{\lambda_{1}}^{0}+n_{\lambda_{2}}^{0}\right) \delta\left(\omega_{\lambda}-\omega_{\lambda_{1}}-\omega_{\lambda_{2}}\right)
    +\left(n_{\lambda_{1}}^{0}-n_{\lambda_{2}}^{0}\right) \delta\left(\omega_{\lambda}+\omega_{\lambda_{1}}-\omega_{\lambda_{2}}\right)
    \right].
\label{SE3}
\end{equation}

\begin{equation}
\begin{aligned}
    \frac{1}{\tau_{\lambda}^{4 \mathrm{ph}}} &= 2 \pi \sum_{\lambda_{1} \lambda_{2} \lambda_{3}}\left|V_{\lambda \lambda_{1} \lambda_{2} \lambda_{3}}\right|^{2} \\
    &\quad \times \left[\begin{array}{l}
    \frac{1}{2} \frac{n_{\lambda_{1}}^{0} n_{\lambda_{2}}^{0}\left(n_{\lambda_{3}}^{0}+1\right)}{\left(n_{\lambda}^{0}+1\right)} \delta\left(\omega_{\lambda}+\omega_{\lambda_{1}}+\omega_{\lambda_{2}}-\omega_{\lambda_{3}}\right) \\
    +\frac{1}{2} \frac{\left(n_{\lambda_{1}}^{0}+1\right) n_{\lambda_{2}}^{0} n_{\lambda_{3}}^{0}}{n_{\lambda}^{0}} \delta\left(\omega_{\lambda}+\omega_{\lambda_{1}}-\omega_{\lambda_{2}}-\omega_{\lambda_{3}}\right) \\
    +\frac{1}{6} \frac{n_{\lambda_{1}}^{0} n_{\lambda_{2}}^{0} n_{\lambda_{3}}^{0}}{n_{\lambda}^{0}} \delta\left(\omega_{\lambda}-\omega_{\lambda_{1}}-\omega_{\lambda_{2}}-\omega_{\lambda_{3}}\right)
    \end{array}\right],
\end{aligned}
\label{SE4}
\end{equation}
where, $V_{\lambda \lambda_{1} \lambda_{2}}$ and $V_{\lambda \lambda_{1} \lambda_{2} \lambda_{3}}$ denote the three-phonon and four-phonon scattering matrix element. The detailed expressions for them can be found in Ref. \cite{feng2016quantum}. $\delta$ is the Dirac delta function which ensures the conservation of energy during the scattering processes. The phonon-isotope scattering can be calculated based on the Tamura theory and the details can be found in Ref. \cite{tamura1983isotope}. The phonon-electron and phonon-hole scattering rates are related to the imaginary part of the phonon self-energy, which can be expressed as,\cite{ponce2016epw}
\begin{equation}
    \frac{1}{\tau_{\lambda}^{\text{ph-el/hole}}} = -\frac{2 \pi}{\hbar} \sum_{m n, \mathbf{k}}\left|g_{m n}^{v}(\mathbf{k}, \mathbf{q})\right|^{2}
    \left(f_{n \mathbf{k}}-f_{m \mathbf{k}+\mathbf{q}}\right) \times \delta\left(\varepsilon_{m \mathbf{k}+\mathbf{q}}-\varepsilon_{n \mathbf{q}}-\hbar \omega_{\lambda}\right)
    \label{SE5}
\end{equation}
where $g_{m n, v}(\mathbf{k}, \mathbf{q})$ quantifies the scattering processes between the electronic state $n \mathbf{k}$ and $m \mathbf{k}+\mathbf{q}$, $f$ is the Fermi-Dirac distribution function, $\varepsilon$ is the electron energy, and $\varepsilon_{F}$ is the Fermi energy determined by the charge carrier concentration.

\section{Band structure of 2D-GaN with and without spin-orbit coupling (SOC)}
\begin{figure}[H]
    \centering
    \includegraphics[width=0.52\columnwidth]{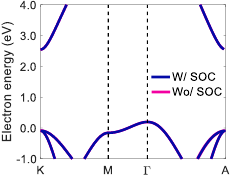}
    \caption{Band structure of 2D-GaN with and without SOC. The electron energy is normalized to the valence band maximum}
    \label{fig: Figure S1}
\end{figure}

Figure \ref{fig: Figure S1} shows that the SOC has almost no effect on the electron band structure in the vicinity of conduction band minimum and valence band maximum. Therefore, the effect of SOC is neglected in our first-principles calculations.

\section{Convergence tests on phonon dispersion}
\begin{figure}[H]
    \centering
    \includegraphics[width=0.60\columnwidth]{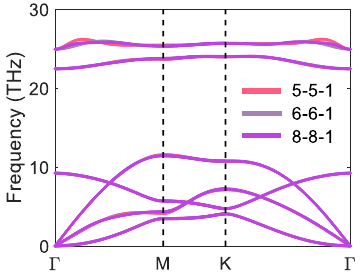}
    \caption{Convergence tests on the phonon dispersion for 2D-GaN with $\mathbf{q}$-point meshes.}
    \label{fig: Figure S2}
\end{figure}

The convergence tests on the phonon dispersion of 2D-GaN along the high-symmetry path are shown in Figure \ref{fig: Figure S2}. It can be seen that the phonon dispersions do not have significant variations using 6 × 6 × 1 and 8 × 8 × 1 $\mathbf{q}$-point meshes. Therefore, we adopt a 6 × 6 × 1 $\mathbf{q}$-point mesh in all our calculations.

\section{Convergence tests on third-order interatomic force constants}
\begin{figure}[H]
    \centering
    \includegraphics[width=0.60\columnwidth]{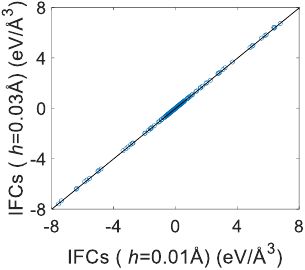}
    \caption{The comparison of 3rd-IFCs calculated by using finite difference method with \textit{h} of 0.01 and 0.03 Å.}
    \label{fig: Figure S3}
\end{figure}

Figure \ref{fig: Figure S3} shows the third-order interatomic force constants (3rd-IFCs) of 2D-GaN calculated using the finite difference method with \textit{h} values of 0.01 and 0.03 Å. It is found that the 3rd-IFCs are insensitive to atomic displacement.

\section{Convergence tests on lattice thermal conductivity}
\begin{figure}[H]
    \centering
    \includegraphics[width=1.00\columnwidth]{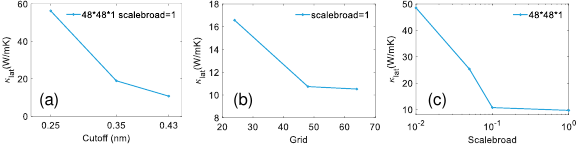}
    \caption{Convergence tests on the lattice thermal conductivity with respect to (a) cutoff distance, (b) $\mathbf{q}$-point meshes, and (c) scalebroad.}
    \label{fig: Figure S4}
\end{figure}

Figure \ref{fig: Figure S4} (a) shows that a cutoff distance of 0.43 nm for the atomic neighborhood is enough to ensure convergence. Additionally, we use the scalebroad=1 to check the convergence with respect to $\mathbf{q}$-point meshes in PBTE and find that the lattice thermal conductivity can be converged with $\mathbf{q}$-point mesh of 48×48×1, as shown in Figure \ref{fig: Figure S4} (b). Then, we fix the $\mathbf{q}$-point mesh as 48×48×1 and conduct the convergence on the scalebroad. We find that the scalebroad should be larger than 0.1, as shown in Figure \ref{fig: Figure S4}(c).

\section{Four-phonon scattering rates and lattice thermal conductivities from RTA and iterative calculation schemes}
\begin{figure}[H]
    \centering
    \includegraphics[width=1.00\columnwidth]{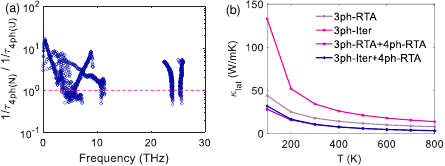}
    \caption{(a) The ratio of four-phonon scattering rates of N and U processes. (b) The lattice thermal conductivity as a function of temperature for three-phonon (RTA), three-phonon (iterative), three-phonon (RTA) + four-phonon (RTA), and three-phonon (iterative) + four-phonon (RTA).}
    \label{fig: Figure S5}
\end{figure}

Figure \ref{fig: Figure S5} shows that the four-phonon scattering rates corresponding to the normal (N) processes are significantly larger than those corresponding to the Umklapp (U) processes. This result suggests that considering the four-phonon scattering at the relaxation time approximation level may be insufficient in calculating the lattice thermal conductivity of 2D-GaN. However, the iterative solutions play a crucial role in determining the three-phonon thermal conductivity. With the four-phonon RTA scattering rate further included, the effect of iterative solutions becomes negligible as shown in Figure \ref{fig: Figure S5}(b). Therefore, it can be expected the iterative solution can also have a weak effect on the lattice thermal conductivity of 2D-GaN. 

\section{Band structure of 2D-GaN and Fermi energy related to the carrier concentrations}
\begin{figure}[H]
    \centering
    \includegraphics[width=0.52\columnwidth]{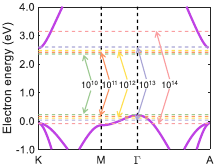}
    \caption{Electron band structure of 2D-GaN along the high symmetry paths. The horizontal lines are Fermi energy related to the carrier concentrations of $10^{10}$ (green), $10^{11}$ (orange), $10^{12}$ (yellow), $10^{13}$ (light purple), and $10^{14}$ (light pink) $ \, \text{cm}^{-2}$ at room temperature. The electron energy is normalized to the valence band maximum.}
    \label{fig: Figure S6}
\end{figure}

Figure \ref{fig: Figure S6} shows the electron band structure of 2D-GaN. The Fermi energy corresponding to different carrier concentrations is represented by horizontal dash lines.

\section{Accumulated lattice thermal conductivity}
\begin{figure}[H]
    \centering
    \includegraphics[width=0.52\columnwidth]{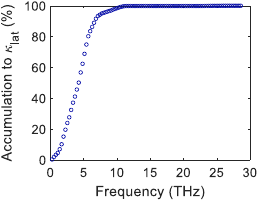}
    \caption{Accumulated lattice thermal conductivity as a function of the frequency at room temperature.}
    \label{fig: Figure S7}
\end{figure}

In Figure \ref{fig: Figure S7}, we show the spectral distribution of the lattice thermal conductivity of 2D-GaN at room temperature. The phonons within the low-frequency range (below 12 THz) have almost all the contribution to the lattice thermal conductivity.

\section{Electron density of state}
\begin{figure}[H]
    \centering
    \includegraphics[width=0.52\columnwidth]{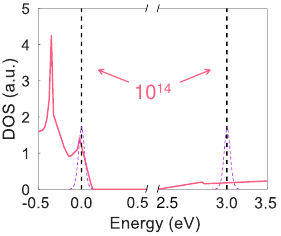}
    \caption{Electron density of state near the valence- and conduction-band edges. The purple dotted curve represents the Fermi window. The electron energy is normalized to the VBM. The position of the Fermi energy for electron and hole concentrations of $10^{14} \, \text{cm}^{-2}$ is indicated with black dashed lines.}
    \label{fig: Figure S8}
\end{figure}

The electron density of states within the Fermi window for \textit{p}-type 2D-GaN is notably larger than that of for \textit{n}-type 2D-GaN at the charge carrier concentration of $10^{14} \, \text{cm}^{-2}$, as shown in Figure \ref{fig: Figure S8}.

\bibliography{bibliography}